\documentclass[11pt,a4paper]{article}
\usepackage{epsfig,graphicx,amsfonts,amsmath}

\RequirePackage{mathrsfs}

\newlength{\dinwidth}
\newlength{\dinmargin}
\newcommand{\Le}{\left(}
\newcommand{\Ra}{\right)}

\newcommand{\beq}{\begin{equation}}
\newcommand{\eeq}{\end{equation}}
\newcommand{\beqar}{\begin{eqnarray}}
\newcommand{\eeqar}{\end{eqnarray}}
\date{}
\begin{document}

\title {{~}\\
{\Large \bf  Abelian Fock - Schwinger representation for the quark propagator in external gauge field}\\}
%

\author{
{~}\\
{\large
M.~A.~Zubkov $^{(1) }$\footnote{On leave of absence from NRC "Kurchatov Institute" - ITEP, B. Cheremushkinskaya 25, Moscow, 117259, Russia}
}\\[7mm]
{\it\normalsize  $^{(1) }$ Physics Department, Ariel University, Ariel 40700, Israel}\\
}

\maketitle
\thispagestyle{empty}

\begin{abstract}
We develop the Diakonov - Petrov approach to the representation of the non - Abelian Wilson loop in an Abelian form. First of all, we extend this approach to the Abelian representation of the parallel transporter corresponding to the open curve rather than to the closed contour. Next, we extend this representation to the case of the non - compact groups. The obtained expressions allow us to derive the Abelian form of the Schwinger - Fock representation for the quark propagator in an external color gauge field.
\end{abstract}

\section{Introduction}

In the gauge theory the basic quantity is the parallel transporter along a curve in space - time:
\begin{equation}
W_q[{ C}_{xy}] =  {\rm P} {\rm exp} \,(i \int_{{ C}_{xy}} A_{\mu} dx^{\mu})\label{W1}
\end{equation}
Here ${ C}_{xy}$ is the curve connecting space - time points $x$ and $y$, $A_\mu$ is the gauge field taken in the representation $q$ of the Lie algebra of the gauge group, while the symbol ${\rm P}$ means ordering along the curve.
The Abelian gauge theories are in many aspects simpler than the non - Abelian ones. It is much more easy to calculate the  parallel transporters (and the other quantities derived from them) within the Abelian gauge theory. That is why one of the most powerful methods of the analytical investigations in strongly coupled gauge theories is the method of Abelian projection \cite{tH}. In this method the dynamics of the non - Abelian gauge theory is expressed in terms of certain Abelian variables obtained as a result of the "projection" of the original non - Abelian ones \cite{P}.

In various applications of the Abelian projection method it is useful to have an Abelian - like representation for the parallel transporter of Eq. (\ref{W1}). The representation of such kind for the Wilson loop (i.e. for the trace of the parallel transporter along the closed curve) has been proposed by D. Diakonov and V. Petrov \cite{DP0,DP,DP1} (see also \cite{KT}). In \cite{Z2002} this representation has been presented in the following form (for the case of the
irreducible representation $q$ of $G=SU(N)$):
\begin{equation}
{\rm Tr}\,{\rm P}\, {\rm exp}\Bigl( i \int_{C_{xy}} A(z) dz \Bigr) =  \int \,  D \mu_{{\cal C},q}(g) \, {\rm exp} \Bigl(i \int_{C_{xy}} \langle 0| A^g(z) |0\rangle dz \Bigr)\, \label{DPf0}
\end{equation}
Here the integration is over the field $g$ that belongs to the gauge group $G$. $A^g$ is the corresponding gauge transformation of the field $A$.  By $|0\rangle$ we denote the eldest vector of the given irreducible representation.
$D\mu_{{\cal C},q}(g) $ is the integration measure over the gauge transformations $g$. It has the form of the product
$$
D \mu_{{\cal C},q}(g)  = e^{\kappa |C_{xy}|} \, Dg
$$
where $Dg$ is the Haar measure on the group that is normalized in the conventional way: at any given point $x$ the integral $\int dg(x)$ is equal to unity.
In Euclidean space - time by $|C_{xy}|$ we denote the length of the path while $\kappa$  depends on the cutoff in the ultraviolet regularized theory. For the case of Minkowski space - time the expression $\kappa |C_{xy}|$  should have the different meaning. Namely, as we will see below, this combination is equal to the product of dimensionless parameter ${\rm log}\, d$ and the number of the points in the discretization of the path $C_{xy}$. After the regularization is taken off in both cases the product $\kappa |C_{xy}|$ becomes divergent, and is to be absorbed by the renormalized quantities. This divergency was not discussed in sufficient details in the original papers by Diakonov and Petrov \cite{DP0,DP}, which became the source of the misunderstanding - in \cite{Iv} it was argued that the Diakonov - Petrov formula of Eq. (\ref{DPf0}) is not valid at all. This misunderstanding has been clarified in \cite{Z2002} devoted to the derivation of Eq. (\ref{DPf0}) in lattice regularization \cite{Creutz}, where the ultraviolet divergent nature of  constant $\kappa$ is especially transparent.


The path integral over $g$ entering Eq. (\ref{DPf0}) reveals an analogy with the
Fock-Feynman-Schwinger representation (FSR) of the fermion propagator. This representation is based on the
Fock-Schwinger proper time  and the Feynman path integral
formalism\cite{1,2}. Various applications of this representation to QED have been used (see, for example, \cite{3}). The applications to QCD have been discussed in
 \cite{4}. The alternative derivation of the FSR  using the stochastic background method has been given in  \cite{5}. The review of the FSR  may be found in \cite{6}. Certain modification of the FSR representation was suggested in Ref. \cite{7}. The application of the mentioned form of the fermion propagator to the  one-loop perturbative amplitudes is especially convenient \cite{8,9}. These calculations are used for the calculation of the effective actions both in QED and in QCD \cite{10}.
The applications of this formalism to the finite temperature theory have been discussed in \cite{11,12,13}.

The FSR representation reduces various physical quantities to the integrals of the
 Wilson loops. This gives one more link of this formalism to the mentioned above Diakonov - Petrov representation of Eq. (\ref{DPf0}). This approach allows to avoid summation of the Feynman diagrams \cite{14,15} for the calculation of various form factors in QCD. The nonperturbative phenomena may be also treated directly with the aid of the FSR \cite{16,17}. For some other features of the FSR we refer to \cite{18,19,20,21,22,23,24,25}.
For the review of the applications of the FSR we refer to \cite {26,Simonov:2002zj}.


In the present paper we establish links between the FSR and the Diakonov - Petrov representation of the Wilson loop of Eq. (\ref{DPf0}).
For the definiteness we consider the quark propagator, but the method developed here works equally well for the fermions from various representations of any compact gauge group.
The Schwinger - Fock representation (SFR) of the fermion propagator in the presence  of external gauge field reduces it to the integral over the trajectories of the fermion. Each term in this integral contains the path ordering exponent of a matrix acting on both Dirac spinor indices and the color ones. This $P$ - ordered exponent may be considered as the parallel transporter corresponding to the composite $G=SU(3) \otimes SO(3,1)$ gauge field (or the $G=SU(3) \otimes SO(4)$ gauge field for the case of Euclidean space - time). We develop further the approach of Diakonov and Petrov in order to represent this $P$ - exponent in the form of an integral over $G$ of the parallel transporter corresponding to a certain Abelian gauge field. As it was mentioned above, this approach was proposed originally for the representation of the non - Abelian Wilson loop in an Abelian form. We develop it further in order to give the Abelian representation of the parallel transporter corresponding to the open curve rather than to the closed contour. Moreover, we extend this approach to the non - compact gauge groups. The obtained Abelian representation of the non - Abelian parallel transporter, being in itself the separate result, allows us to obtain the desired Abelian form of the Schwinger - Fock representation for the fermion propagator, in which both color and spin $P$ - exponents are represented in the form of the parallel transporter of an Abelian gauge field.

The paper is organized as follows. In Sect. \ref{SectDP} we propose the Diakonov - Petrov like expression for the parallel transported along the open curve. In Sect. \ref{SectProp} we consider the theory in Minkowski space - time and apply the obtained expression for the derivation of the Abelian path integral representation of quark propagator (that gives the Abelian version of the FSR). In Sect. \ref{SectProp2} we derive the similar representation in Euclidean space - time. In Sect. \ref{SectConcl} we end with the conclusions.

\section{Diakonov - Petrov like expression for the parallel transporter}

\label{SectDP}

The original formulation of the Diakonov - Petrov representation was given for the Wilson loop corresponding to a compact gauge group and the closed contour. Below we derive the following representation for the parallel transporter along the path $C_{xy}$:
\begin{equation}
{\rm P}\, {\rm exp}\Bigl( i \int_{C_{xy}} A(z) dz \Bigr) = e^{\kappa |C_{xy}|} \, \int \, g_x |0\rangle \, Dg \, {\rm exp} \Bigl(i \int_{C_{xy}} \langle 0| A^g(z) |0\rangle dz \Bigr)\,\langle 0| g^{-1}_y \label{PTDP}
\end{equation}
Here the integration is over the field $g$ that belongs to the gauge group $G$. Let us require first that the group $G$ is compact.   $Dg$ is the Haar measure on the group that is normalized in the conventional way: at any given point $x$ the integral $\int dg(x)$ is equal to unity. The representation is assumed to be irreducible. As $|0\rangle$ we take an arbitrary fixed vector of the given representation (we may take the eldest vector, for definiteness). The given representation is irreducible, and the integration measure $dg$  is invariant under the transformation $g \to U g$ for any  $U\in G$. Then for any $|0\rangle$  we have
$$
U \int dg g|0\rangle \langle 0|g^{-1} U^{-1}= \int dg g|0\rangle \langle 0|g^{-1}
$$
That means that $\hat M = \int dg g|0\rangle \langle 0|g^{-1}$ commutes with any element of the gauge group. For the irreducible representations such operator is proportional to unity:
\begin{equation}
\int dg g|0\rangle \langle 0|g^{-1} = 1/d\label{1d}
\end{equation}
where $d$ is a constant. As it was mentioned in the introduction, in the case of the irreducible representation of unitary group we may take the eldest vector, while $d$ is the dimension of the given representation. $\kappa$ is a constant that depends on the regularization. It is linear divergent in ultraviolet and is to be absorbed by the other divergent quantities during the procedure of renormalization. In Euclidean space - time by $|C_{xy}|$ we denote the length of the path. For the case of Minkowski space - time the expression $\kappa |C_{xy}|$  should have the different meaning. Namely, as we will see below, this combination is equal to the product of dimensionless parameter ${\rm log}\, d$ and the number of the points in the discretization of the path $C_{xy}$. After the regularization is taken off in both cases the product $\kappa |C_{xy}|$ becomes divergent, and is to be absorbed by the renormalized quantities.

In order to prove Eq. (\ref{PTDP}) let us represent the path as a sequence of small intervals: $C_{xy} = C_{x z_1}\cup C_{z_1 z_2} \cup ... \cup C_{z_{N-1} y}$. Let us identify $x \equiv z_0$, $y=z_N$. Then we may write:
\begin{eqnarray}
{\rm P}\, {\rm exp}\Bigl( i \int_{C_{xy}} A(z) dz \Bigr) & = & \prod_{i=0...N-1}  {\rm exp}\Bigl( i \int_{C_{z_i z_{i+1}}} A(z) dz \Bigr)\nonumber\\ & = &  d^{N+1}\int  g_0 |0\rangle dg_0 \langle 0|g^{-1}_0 \prod_{i=0...N-1}  {\rm exp}\Bigl( i \int_{C_{z_i z_{i+1}}} A(z) dz \Bigr) \int  g_i |0\rangle dg_i \langle 0|g^{-1}_i\nonumber\\ & = & d^{N+1} \int  g_0 |0\rangle dg_0 \langle 0| \prod_{i=0...N-1}  {\rm exp}\Bigl( i \int_{C_{z_i z_{i+1}}} g^{-1}_i A(z) g_i dz \Bigr) \int g^{-1}_i g_{i+1} |0\rangle dg_{i+1} \langle 0|g^{-1}_{i+1}\nonumber\\ & = & d^{N+1} \int  g(x) |0\rangle dg(x)  \prod_{i=0...N-1}  \int \nonumber\\&&{\rm exp}\Bigl( i \int_{C_{z_i z_{i+1}}}\langle 0|( g^{-1}(z) A(z) g(z) dz + g^{-1}(z) d g(z))|0\rangle\Bigr)   dg(z_{i+1}) \langle 0|g^{-1}(z_{i+1})
\nonumber\\ & = & e^{\kappa |C_{xy}|} \int  g(x) |0\rangle Dg\,   {\rm exp}\Bigl( i \int_{C_{x y}}\langle 0|( g^{-1} A(z) g dz -i g^{-1} d g)|0\rangle\Bigr)  \langle 0|g^{-1}(y)
\end{eqnarray}
Thus we identify $\kappa$ with $\frac{{\rm log}\,d}{\Delta z}$ (where $\Lambda = \frac{1}{\Delta z}$ has the meaning of the ultraviolet cutoff for the discretization of the path) and imply $|C_{xy}| = \int_{C_{xy}}\sqrt{dz^2}$ for space - time with Euclidean signature, for Minkowski space - time we consider the product $\kappa |C_{xy}|$ as the product of ${\rm log}\,d$ and the number of points in the discretization of the path. In both cases we   arrive at Eq. (\ref{PTDP}).

For the non - compact group $G$ the situation is more complicated because the integral $\int dg$ is divergent. The non - normalized measure still may be defined, but it leads to the divergent quantity $1/d$ in Eq. (\ref{1d}). Therefore, for the case of the non - compact group we will use the over - completeness of the coherent state system. To be explicit, let us consider the group $SL(2,C)$, which will be used below because it is locally isomorphic to the Lorentz group $SO(3,1)$. We consider the fundamental representation of $SL(2,C)$, in which its elements are parametrized as follows:
$$
g = e^{i a_k \sigma^k}
$$
where $a_k$ is the complex - valued vector while $\bar{a}_k$ is its complex conjugation. The real parts of $a_k$ belong to the interval $[0;2\pi)$ while the imaginary parts may be arbitrary real numbers. We will use the completeness of the subgroup $SU(2)$ of rotations, that are generated by the elements of $SL(2,C)$ with the real - valued $a_k$. Let us denote $g_{SU(2)} \in SU(2) \subset SL(2,C)$. The following property holds:
\begin{equation}
\int dg_{SU(2)} g_{SU(2)}|0\rangle \langle 0|g^{-1}_{SU(2)} = 1/{d_{SU(2)}} = 1/2 \label{1d2}
\end{equation}
where $d g_{SU(2)}$ is the conventionally normalized Haar measure on $SU(2)$.
Thus for the parallel transporter corresponding to $A\in sl(2,C)$ we obtain
\begin{equation}
{\rm P}\, {\rm exp}\Bigl( i \int_{C_{xy}} A(z) dz \Bigr) = e^{\kappa |C_{xy}|} \, \int \, g_x |0\rangle \, D_{SU(2)} g \, {\rm exp} \Bigl(i \int_{C_{xy}} \langle 0| A^g(z) |0\rangle dz \Bigr)\,\langle 0| g^{-1}_y \label{PTDP2}
\end{equation}
Here the integral is over the subgroup $SU(2)\subset SL(2,C)$.

The generalization of this construction to any noncompact gauge group (that contains the compact subgroup, such that it generates the complete system of coherent states) is straightforward.

\section{Path integral representation for the fermion propagator (Minkowski space - time)}

\label{SectProp}

In this section we briefly review the conventional FSR for the fermion propagator, and after that we will propose the Abelian form of the FSR based on the derived above Eq. (\ref{PTDP}). We consider the fermion propagator $S_{xy}$ (dependent on space - time points $x$ and $y$) in the presence of the external gauge field $A$ that may, in principle, belong to any compact gauge group. But for the definiteness we restrict ourselves by the color gauge group $SU(3)$:
\beq\label{1Sec1}
\Le\, i\,\hat{D}\,-\,m\Ra_{x}\,S_{x\,y}\,=\,\delta_{x y}\,,\,\,\,\,D_{x}^{\mu}\,=\,\partial_{x}^{\mu}\,-\,i\, A^{\mu}\,,
\eeq
First let us introduce the function
\beq\label{1Sec10}
\Le\,\hat{D}^{2}\,+\,m^{2}\,\Ra_{x}\,\Delta_{x y}\,=\,-\,\delta_{x y}\,
\eeq
The fermion propagator is expressed through this function as follows:
\beq\label{1Sec110}
S_{x\,y}\,=\,\Le\,i\,\hat{D}\,+\,m\Ra_{x}\,\Delta_{x y}\,=\,\Le\,(i\,\hat{\partial}\,+\,m) + \, \hat{A}\Ra_{x}\,\Delta_{x y}.
\eeq

In order to derive the path integral representation for $\Delta_{x\,y}$ let us represent it as follows:
\begin{equation}
\Delta_{x\, y} = - \langle y | \frac{1}{\hat{\cal H}}| x \rangle
\end{equation}
where
\begin{equation}
\hat{\cal H} = \hat{D}^{2}\,+\,m^{2} = D^2\, - \frac{1}{2}F_{ij}\sigma^{ij} \, + m^2
\end{equation}
where $F^{ij} = \partial_{[i}A_{j]} - i[A_i,A_j]$ is the field strength while $\sigma^{ij} = \frac{i}{2}[\gamma^i,\gamma^j]$ is the generator of the Lorentz group $SO(3,1)$ written in the $4\times 4$ spinor representation. For the case of quarks, when the gauge field belongs to the $SU(3)$ group, $\cal H$ is the $(3\times 3) \otimes (4\times 4) = (3\times 4)\times (3\times 4)$ matrix that belongs to the representation of the group $SU(3)\otimes SO(3,1)$, which is the product of the fundamental representation of $SU(3)$ and the spinor representation of $SO(3,1)=SL(2,C)/Z_2$.

Assuming that $\cal H$ contains the small imaginary part: $\hat{\cal H} \to \hat{\cal H} - i 0$, we may represent $\Delta$ as follows:
\begin{equation}
\Delta_{x\, y} = i \int_{0}^\infty ds \langle y| e^{- i \hat{\cal H} ds }| x \rangle
\end{equation}
Then the standard quantum mechanical path integral representation for the evolution operator gives
\begin{equation}
\Delta_{x\, y} = i \int_{0}^\infty ds \int_{z(0)=x}^{z(s)=y} Dz P \,e^{ i \int_0^s d\tau \Bigl(\frac{\dot{z}^2(\tau)}{4} - m^2 + A(z(\tau))\dot{z}(\tau) + \frac{1}{2} F_{ij}(z(\tau)) \sigma^{ij} \Bigr) }\label{DELTA}
\end{equation}

The following comments are in order. First of all, the $P$ - ordered exponent entering Eq. (\ref{DELTA}) belongs to the reducible representation of $SU(3)\otimes SL(2,C)$.

However, we may represent it in the block - diagonal form, where the nonzero diagonal blocks belong to the irreducible representations, which are complex conjugate with respect to each other.
The boost generators are
$$
\sigma^{0k}= i \left(\begin{array}{cc} \sigma^k& 0 \\ 0 & - \sigma^k \end{array}\right)
$$
while the rotation generators are
$$
\sigma^{ij}= \epsilon^{ijk} \left(\begin{array}{cc} \sigma^k& 0 \\ 0 &  \sigma^k \end{array}\right)
$$
Therefore, we may write:
\begin{eqnarray}
&& P \,e^{ i \int_0^s d\tau \Bigl( A(z(\tau))\dot{z}(\tau) + \frac{1}{2} F_{ij}(z(\tau)) \sigma^{ij} \Bigr) } =  \\ && \left(\begin{array}{cc} P \,e^{ i \int_0^s d\tau \Bigl( A(z(\tau))\dot{z}(\tau) +  i F_{0j}(z(\tau)) \sigma^{j} + \frac{1}{2} F_{ij}(z(\tau))\epsilon^{0ijk} \sigma^{k} \Bigr) }& 0 \\ 0 & P \,e^{ i \int_0^s d\tau \Bigl( A(z(\tau))\dot{z}(\tau) -  i F_{0j}(z(\tau)) \sigma^{j} + \frac{1}{2} F_{ij}(z(\tau))\epsilon^{0ijk} \sigma^{k} \Bigr) } \end{array}\right)\nonumber
\end{eqnarray}
We are able to apply Eqs. (\ref{PTDP}) and (\ref{PTDP2}) to each of the blocks of this matrix. As a result we obtain
\begin{eqnarray}
&&\Delta_{x\, y} =  i \int_{0}^\infty ds \int_{z(0)=x}^{z(s)=y} Dz  \,e^{ i \int_0^s d\tau \Bigl(\frac{\dot{z}^2(\tau)}{4} - m^2\Bigr) + \kappa |C_{xy}|}\nonumber\\ &&
 \int \,  \, Dg \,D_{SU(2)\subset SL(2,C)}h\, {\rm exp} \Bigl(i \int_{C_{xy}} \langle 0| \Big[g^{+}(z)A(z) g(z) dz +  \frac{1}{2} g^+(z) F_{ij}(z(\tau)) g(z) h^{-1}(z)\sigma^{ij}h(z) d \tau \Big]|0\rangle  \Bigr)\nonumber\\ &&
 {\rm exp} \Bigl( \int_{C_{xy}} \langle 0| \Big[g^{+}(z)d g(z) + h^{-1}(z)d h(z) \Big]|0\rangle  \Bigr)\nonumber\\ &&h_x \, g_x |0\rangle\,\langle 0| g^+_y \, h^{-1}_y\label{ASFRM}
\end{eqnarray}
Here $g\in SU(3)$, while $h \in SU(2)\subset SL(2,C)$, $|0\rangle = |0\rangle_{SU(3)} \otimes|0\rangle_{SL(2,C)}$. Dealing with the quarks we choose, say, $|0\rangle_{SU(3)} = \left( \begin{array}{c} 1 \\ 0 \\ 0 \end{array} \right)$ and $|0\rangle_{ SL(2,C)} = \left( \begin{array}{c} 1 \\ 0 \\ 1 \\ 0 \end{array} \right)$.

We may parametrize matrix $h \in SL(2,C)$ as follows:
$$
h = \left(\begin{array}{cc} e^{i a_k \sigma^k}& 0 \\ 0 & e^{i \bar{a}_k \sigma^k} \end{array}\right)
$$
where $a_k$ is the complex - valued vector while $\bar{a}_k$ is its complex conjugation. The real parts of $a_k$ belong to the interval $[0;2\pi)$ while the imaginary parts may be arbitrary real numbers. For the subgroup $SU(2)\subset SL(2,C)$ entering Eq. (\ref{ASFRM}) we have $a_k \in R$.

\section{Path integral representation for the fermion propagator (Euclidean space - time)}

\label{SectProp2}

In the case of Euclidean propagator we express $\hat{D} = D_k \Gamma^k$ through the Euclidean gamma - matrices: $\Gamma^4 = \gamma^0$, and $\Gamma^k=i \gamma^k$ (for $k=1,2,3$). Now we have:
\beq\label{1Sec112}
S_{x\,y}\,=\,\Le\,-i \,\hat{D}\,+\,m\Ra_{x}\,\Delta_{x y}\,=\,\Le\,(-i\,\hat{\partial}\,+\,m) - \, \hat{A}\Ra_{x}\,\Delta_{x y}.
\eeq
with
\begin{equation}
\Delta_{x\, y} =  -\langle y | \frac{1}{\hat{\cal H}}| x \rangle
\end{equation}
where
\begin{equation}
\hat{\cal H} = -\hat{D}^{2}\,+\,m^{2} = - D^2\, + \frac{1}{2}F_{ij}\Sigma^{ij} \, + m^2
\end{equation}
where $F^{ij} = \partial_{[i}A_{j]} - i[A_i,A_j]$ is the field strength while $\Sigma^{ij} = \frac{i}{2}[\Gamma^i,\Gamma^j]$ is the generator of the group $SO(4)=SU(2)\otimes SU(2)/Z_2$ written in the $4\times 4$ spinor representation. Thus, $\cal H$ here is the $(3\times 3) \otimes (4\times 4) = (3\times 4)\times (3\times 4)$ matrix that belongs to the representation of the group $SU(3)\otimes SO(4)$, which is the product of the fundamental representation of $SU(3)$ and the spinor representation of $SO(4)=SU(2) \otimes SU(2)/Z_2$.

We may represent $\Delta$ as follows:
\begin{equation}
\Delta_{x\, y} =  \int_{0}^\infty ds \langle y| e^{-  \hat{\cal H} ds }| x \rangle
\end{equation}
In Euclidean space the standard quantum mechanical path integral representation for the evolution operator gives
\begin{equation}
\Delta_{x\, y} = \int_{0}^\infty ds \int_{z(0)=x}^{z(s)=y} Dz P \,e^{ -\int_0^s d\tau \Bigl(\frac{\dot{z}^2(\tau)}{4} + m^2 - i A(z(\tau))\dot{z}(\tau) + \frac{1}{2} F_{ij}(z(\tau)) \Sigma^{ij} \Bigr) }
\end{equation}
Again, the $P$ - exponent has the block - diagonal form, but it belongs to the  representation of $SU(3) \otimes SU(2)_R \otimes SU(2)_L$. The upper block belongs to the fundamental representation of $SU(3) \otimes SU(2)_R$ while the lower block belongs to the fundamental representation of $SU(3) \otimes SU(2)_L$. As above, we apply Eq. (\ref{PTDP}) to each of the blocks and obtain:
\begin{eqnarray}
\Delta_{x\, y} & = & \int_{0}^\infty ds \int_{z(0)=x}^{z(s)=y} Dz  \,e^{ - \int_0^s d\tau \Bigl(\frac{\dot{z}^2(\tau)}{4} + m^2\Bigr) + \kappa \int_0^s \sqrt{\dot{z}^2(\tau)}d\tau}\nonumber\\ &&
 \int \,  \, Dg \,Dh\, {\rm exp} \Bigl(i \int_{C_{xy}} \langle 0| \Big[g^{+}(z)A(z) g(z) dz +  \frac{i}{2} g^+(z) F_{ij}(z(\tau)) g(z) h^{-1}(z)\Sigma^{ij}h(z) d \tau \Big]|0\rangle  \Bigr)\nonumber\\ &&
 {\rm exp} \Bigl( \int_{C_{xy}} \langle 0| \Big[g^{+}(z)d g(z) + h^{+}(z)d h(z) \Big]|0\rangle  \Bigr)\nonumber\\ &&h_x \, g_x |0\rangle\,\langle 0| g^+_y \, h^{+}_y\label{DE}
\end{eqnarray}
Here $g\in SU(3)$, while $h \in SU(2)\otimes SU(2)$, $|0\rangle = |0\rangle_{SU(3)} \otimes|0\rangle_{SU(2)\otimes SU(2)}$. Dealing with quarks we choose,  $|0\rangle_{SU(3)} = \left( \begin{array}{c} 1 \\ 0 \\ 0 \end{array} \right)$ and $|0\rangle_{SU(2)\otimes SU(2)} = \left( \begin{array}{c} 1 \\ 0 \\ 1 \\ 0 \end{array} \right)$.

Now we may parametrize matrix $h$ as follows:
$$
h = \left(\begin{array}{cc} e^{i a_k \sigma^k}& 0 \\ 0 & e^{i {b}_k \sigma^k} \end{array}\right)
$$
where $a_k$ and $b_k$ are the real - valued vectors. Their components belong to the interval $[0;2\pi)$.

It is worth mentioning that the physical meaning of constant $\kappa$ may be seen through analysis of the induced kinetic part of the quark action in the Schwinger  picture:
 $$
 K =  \int_0^{s} d\tau \Bigl(\frac{\dot{z}^2(\tau)}{4} + m^2\Bigr) - \kappa \int_0^{s} \sqrt{\dot{z}^2(\tau)}d\tau
 $$
  Up to the reparametrizations this action is equivalent to  
 $$
 K =  \int_0^{1} d\tau \Bigl(\frac{\dot{z}^2(\tau)}{4e(\tau)} + e(\tau) m^2\Bigr) - \kappa \int_0^{1} \sqrt{\dot{z}^2(\tau)}d\tau
 $$
 with varying one - dimensional metric $e(\tau)$ and the constraint $\int_0^1 e(\tau) d\tau = s$. 
 Minimum of this expression as a functional of $e(\tau)$ gives  classical equations of motion
 $$
 -\frac{\dot{z}^2(\tau)}{4e^2(\tau)} + m^2 =0
 $$ 
 Solution of this equation for $e(\tau)$ being substituted to the action gives:
 $$
 K = (m - \kappa) \int_0^{1} \sqrt{\dot{z}^2(\tau)}d\tau
 $$
 One can recognize in this expression the physical meaning of constant $\kappa$ - it is the contribution to the renormalized quark mass, at least, on the level of classical equations of motion. 

\section{An Abelian version of the Field Correlators Method with applications to QCD }

Method of field correlators (FCM) has been developed by Yu. A. Simonov and co - authors for the non - perturbative treatment of QCD (see, for example, \cite{Simonov:2002zj}). Its basis is the set of correlators of the non - Abelian field strength $F$. A lot of efforts have been invested in order to define the gauge invariant combinations that are to substitute the original non - invariant field strength. For that purpose a reference point $X$ in space - time is to be fixed, and instead of the field strength the following combination has been defined:
$$
\hat{F}_{\mu \nu} = \Phi(X,x) F_{\mu \nu} \Phi(x,X)
$$ 
where 
$$
\Phi(X,x) = P\, {\rm exp}\,\Big(i\int^x_X A(z)dz \Big)
$$
integral here is along the path that connects reference point $X$ with the given point $x$, where the field strength is located. The resulting formalism depends explicitly on the reference point $X$ as well as on the form of the paths connecting points $X$ and $x$. Below we propose version of this formalism that is based on the Abelian representation of Wilson lines and propagators, and is free from the mentioned above unphysical dependence on reference point $X$.

For definiteness using Eq. (\ref{ASFRMc}) we represent  meson propagator $G_M(x,y)$ through the integral over the closed paths that contain points $x$ and $y$. Each term in this integral is given by the integral over $h$ and $g$ of an expression that contains the exponent of an integral over the contour $C$ present in Eq. (\ref{DE}). Since the contour $C$ in this expression is closed, we may transform the integral over this path using Stokes theorem to the integral over the two - dimensional surface bounded by $C$ (see Eq. (47) of \cite{Simonov:2002zj}). In quenched approximation in Euclidean space - time we get
\begin{eqnarray}
G_M(x,y) &=& \Big\langle {\rm tr} \,\int_0^\infty ds_1 \int_0^\infty ds_2 \int Dg Dh \, e^{-K_1 - K_2}Dz_1 Dz_2 \langle 0 | g_y^+ h_y^{-1}\Gamma^{(f)} (m_1-i\hat{D})h_x \, g_x |0\rangle\,\nonumber\\&&\langle 0| g^+_x h^{-1}_x \Gamma^{(i)} (m_2-i\hat{D})h_y \, g_y |0\rangle\, {\rm exp} \Bigl(i \int_{\partial {\cal A}= C} {\cal F}[g,h,A]  \Bigr) \Big\rangle_A\label{GM}
\end{eqnarray}
with $K_i =  \int_0^{s_i} d\tau \Bigl(\frac{\dot{z}^2(\tau)}{4} + m^2\Bigr) - \kappa \int_0^{s_i} \sqrt{\dot{z}^2(\tau)}d\tau$. $\Gamma^{(i)}$ and $\Gamma^{(f)}$ are the matrices specific for the initial and final states of mesons, $m_i$ are the masses of two quarks, $\langle ... \rangle_A$ is averaging with respect to the color gauge field $A$. The integral of $\cal F$ is over the area of the surface bounded by the curve $C$ consisted of the two paths connecting points $x$ and $y$ represented by variables $z_1$ and $z_2$.  Emergent curvature $\cal F$ entering this expression has the form:
\begin{eqnarray}
{\cal F}  &=& d \langle 0| \Big[g^{+}(z)A(z) g(z) dz +  \frac{1}{2} g^+(z) F_{ij}(z(\tau)) g(z) h^{-1}(z)\Sigma^{ij}h(z) d \tau \nonumber\\ && + g^{+}(z)d g(z) + h^{-1}(z)d h(z) \Big]|0\rangle
\end{eqnarray}
The presence of the virtual quark loops will modify this expression by the multiplication (under the averaging over $A$) of the above expression by the sum over the virtual quark loops. The term corresponding to each virtual quark loop is given by an expression similar to that of Eq. (\ref{GM}) (in these expressions certain modifications are needed, in particular, $\int ds$ is to be substituted by $\int ds/s$, see, e.g. Eq. (134) in \cite{Simonov:2002zj}).
The given form of the meson propagator may be useful for the analysis of various non - perturbative properties of heavy mesons including those related intimately to confinement.
The Abelian representation of the spin term in the FSR may simplify somehow the consideration of the spin depending effects.

Let us recall that the functional integral over $A$  (both action and integration measure) is gauge invariant. Therefore, we may actually fix the gauge and arrive at 
\begin{eqnarray}
G_M(x,y) &=& \Big\langle {\rm tr} \,\int_0^\infty ds_1 \int_0^\infty ds_2 \int Dh \, e^{-K_1 - K_2}Dz_1 Dz_2 \langle 0 | h_y^{-1}\Gamma^{(f)} (m_1-i\hat{D})h_x \, |0\rangle\,\nonumber\\&&\langle 0|  h^{-1}_x \Gamma^{(i)} (m_2-i\hat{D})h_y \, |0\rangle\, {\rm exp} \Bigl(i \int_{\partial {\cal A}= C} {\cal F}[h,A]  \Bigr) \Big\rangle_A\label{GM}
\end{eqnarray}
with 
\begin{eqnarray}
{\cal F}  &=& d \langle 0| \Big[A(z) dz +  \frac{1}{2} F_{ij}(z(\tau)) h^{-1}(z)\Sigma^{ij}h(z) d \tau \nonumber\\ &&  + h^{-1}(z)d h(z) \Big]|0\rangle
\end{eqnarray}
In this expression we neglect configurations of paths $z(\tau)$ with self - intersections. Notice, that in QCD the integration variable $h$ cannot be removed in a similar way. The above expression is already not gauge invariant, which means that we have partially fixed the specific gauge corresponding to the given meson propagator. The choice 
$|0\rangle_{SU(3)} = \left( \begin{array}{c} 1 \\ 0 \\ 0 \end{array} \right)$ gives us
\begin{eqnarray}
{\cal F}  &=& d A^{11}(z) dz +  \frac{1}{2} d F^{11}_{ij}(z(\tau))\langle 0| h^{-1}(z)\Sigma^{ij}h(z) |0\rangle d \tau \nonumber\\ &&  -\langle 0|h^{-1} dh \wedge h^{-1}d h |0\rangle
\end{eqnarray}
where 
$|0\rangle$ corresponds to $SO(4)$ only. Now we choose  $|0\rangle_{ SO(4)} = \left( \begin{array}{c} 1 \\ 0 \\ 1 \\ 0 \end{array} \right)$ and represent 
$$
h = \left(\begin{array}{cc} e^{i a_k \sigma^k}& 0 \\ 0 & e^{i {b}_k \sigma^k} \end{array}\right)=\left(\begin{array}{cc} h_1& 0 \\ 0 & h_2  \end{array}\right)
$$
where $h_i \in SU(2)$.
We obtain for the effective "Abelian field strenfgth":
\begin{eqnarray}
{\cal F}  &=& d A^{11}(z) \wedge dz +  \frac{1}{2} d \Big(F^{11}_{ij}(z) \Sigma_h^{ij}\Big) \wedge d \tau - \sum_i \Big[h_i^{+} dh_i \wedge h_i^{+}d h_i\Big]^{11} \label{AF}
\end{eqnarray}
Here 
$$
\Sigma_h^{ij}(z) = \epsilon^{ijk}\sum_{a=1,2} \Big[h^+_a(z) \sigma^k h_a(z)\Big]^{11}, \quad i,j = 1,2,3
$$
and
$$
\Sigma_h^{4j}(z) = \sum_{a=1,2} (-1)^{a}\Big[h^+_a(z) \sigma^j h_a(z)\Big]^{11}, \quad j = 1,2,3
$$

In order to consider properties of confining vacuum of QCD we neglect pre - exponential factors and  assume the so - called Gaussian dominance. This gives for meson propagator
\begin{eqnarray}
G_M(x,y) &\sim &  \,\int_0^\infty ds_1 \int_0^\infty ds_2 \int Dh \, e^{-K_1 - K_2}Dz_1 Dz_2 \, {\rm exp} \Bigl(-\frac{1}{2}\int \int \langle \langle {\cal F} {\cal F}\rangle \rangle  \Bigr) \label{GM}
\end{eqnarray}
Here the integrals in exponent are over the surfaces of minimal area spanned on the worldlines of quarks.

The standard simplification is consideration of heavy quark mesons and neglecting contributions of spin. For the evaluation of effective static inter - quark potential we restrict ourselves by the world - trajectories of quarks that correspond to their static positions.   The dominant contribution to Eq. (\ref{GM}) comes from the cumulant of the Abelian part of the $SU(3)$ field strength. Symmetry fixes the form of the leading cumulant (see \cite{Simonov}):
\begin{eqnarray}
&& \langle \langle \partial_{[\mu_1} A^{11}_{\nu_1]}(z_1) \partial_{[\mu_2} A^{11}_{\nu_2]}(z_2) \rangle \rangle = N_c (\delta_{\mu_1 \mu_2}\delta_{\nu_1  \nu_2} - \delta_{\mu_1 \nu_2}\delta_{\nu_1  \mu_2}) D(z_1-z_2) \nonumber\\&& + \frac{N_c}{2} \Big(\frac{\partial}{\partial z^{\mu_1}_{1}}(z^{\mu_2}_{ 2} \delta_{\nu_1 \nu_2} - z^{\nu_2}_{ 2} \delta_{\nu_1 \mu_2})+\frac{\partial}{\partial z^{\nu_1}_{ 1}}(z^{\nu_2}_{ 2} \delta_{\mu_1 \mu_2} - z^{\mu_2}_{ 2} \delta_{\mu_1 \nu_2})\Big) D_1(z_1 - z_2)
\end{eqnarray}
Functions $D(z)$ and $D_1(z)$ here differ by definition from their counterparts in the original version of FCM. However, they contain the same information about confinement, and, moreover, the static potential in the considered approximation is expressed through them in the same way as in the standard FCM. In particular, we have for the string tension:
\begin{equation}
\sigma = \frac{1}{2}\int d^2 z D(z) \label{sigma} 
\end{equation}
A reasonable form that fits experimental data for the function $D(z)$ may be given in the form
$$
D(z) = \frac{\sigma}{\pi \lambda^2}e^{-|z|/\lambda}
$$
with  parameter $\lambda \approx 0.2$ fm. It can be shown that function $D_1$ results in the appearance of Coulomb interaction, for more details see \cite{Simonov:2018cbk} and references therein. In principle the parametrization of meson correlator in the form of Eq. (\ref{GM}) with $\cal F$ given by Eq. (\ref{AF}) does not change the actual phenomenological applications of the FCM. However, the use of effective Abelian field strength instead of the nonabelian one represents a certain methodological simplification on the way to this expression. It is also worth mentioning, that functions $D$ and $D_1$ may in principle be calculated using precise methods of lattice numerical simulations. Their forms are to be different from those of their non - Abelian counterparts. The two forms may be compared with each other as well as the values of string tension extracted from them according to Eq. (\ref{sigma}). The latter values are to be compared to experiment. The same refers also to the other physical quantities that may be expressed through functions $D$ and $D_1$ (like meson masses etc, again for the details see \cite{Simonov:2018cbk} and references therein).


\section{Conclusions}

\label{SectConcl}

In the present paper we extend the Diakonov - Petrov representation of the Wilson loop of Eq. (\ref{DPf0}) to the open path, and to the non - compact gauge groups. More explicitly, we consider the noncompact $SL(2,C)$ group, the generalization of our construction to the other non - compact groups is straightforward. The derived representation is used to obtain the version of
Fock-Feynman-Schwinger representation (FSR) for the quark propagator in external gauge field in Euclidean space - time:
\beq\label{1Sec112}
S_{x\,y}\,=\,\Le\,-i\,\hat{D}\,+\,m\Ra_{x}\,\Delta_{x y}\,=\,\Le\,(-i\,\hat{\partial}\,+\,m) - \, \hat{A}\Ra_{x}\,\Delta_{x y}.
\eeq
in which $\Delta_{x y}$ is given by the following Abelian representation
\begin{eqnarray}
&&\Delta_{x\, y} =   \int_{0}^\infty ds \int_{z(0)=x}^{z(s)=y} Dz  \,e^{ - \int_0^s d\tau \Bigl(\frac{\dot{z}^2(\tau)}{4} + m^2\Bigr) + \kappa\int_0^s \sqrt{\dot{z}^2(\tau)}d\tau}\nonumber\\ &&
 \int \,  \, Dg \,D_{}h\, {\rm exp} \Bigl(i \int_{C_{xy}} \langle 0| \Big[g^{+}(z)A(z) g(z) dz +  \frac{1}{2} g^+(z) F_{ij}(z(\tau)) g(z) h^{-1}(z)\Sigma^{ij}h(z) d \tau \Big]|0\rangle  \Bigr)\nonumber\\ &&
 {\rm exp} \Bigl( \int_{C_{xy}} \langle 0| \Big[g^{+}(z)d g(z) + h^{-1}(z)d h(z) \Big]|0\rangle  \Bigr)\nonumber\\ &&h_x \, g_x |0\rangle\,\langle 0| g^+_y \, h^{-1}_y\label{ASFRMc}
\end{eqnarray}
Here the integral is over the paths $z(\tau)$ connecting points $x$ and $y$, and over the elements of the gauge group $g\in SU(3)$. The integral $Dh$ is over $h \in SU(2)\otimes SU(2)$.  $|0\rangle = |0\rangle_{SU(3)} \otimes|0\rangle_{SO(4)}$ is a certain vector in the fundamental representation of $SU(3)$ and the 4 - dimensional spinor representation of $SO(4)$. It may be chosen in the form $|0\rangle_{SU(3)} = \left( \begin{array}{c} 1 \\ 0 \\ 0 \end{array} \right)$ and $|0\rangle_{ SO(4)} = \left( \begin{array}{c} 1 \\ 0 \\ 1 \\ 0 \end{array} \right)$.
In Minkowski space - time the representation is similar and is given by Eq. (\ref{ASFRM}).
Constant $\kappa$ is linear divergent in ultraviolet. To be explicit, the path integral representation of Eq. (\ref{ASFRMc}) is to be calculated after a certain ultraviolet regularization is applied (say, we may use the lattice regularization as in \cite{Z2002}). The fermion propagator in external gauge field enters various expressions for the physical observables. In these expressions the integrals over $g$ and $h$ are the sources of the ultraviolet dicergencies \cite{Iv} to be cancelled by the ultraviolet divergent constants $\kappa$ (see \cite{Z2002} for the details).

We expect that the proposed above version of the FSR may be applied to the analytical study of various properties of QCD within various phenomenological models. The field correlator method \cite{Simonov:2018cbk} may be modified to include the Abelian - like representation given above. In particular, within the approach of \cite{Simonov:2002zj} in the expression for the propagators of mesons the FSR for the fermion propagators are combined to give the integral over the closed loops. We have shown how an Abelian version of the field correlator method (FCM) allows to calculate basic quantities like the string tension. Both versions of the FCM may be compared, in principle, using lattice numerical simulations. Those simulations should give explicit expressions for basic functions $D$ and $D_1$ of both approaches. These functions and physical quantities expressed through them  may be compared to experiment.  

We also expect that the proposed Abelian representation for the quark propagator may allow to replace in various physical applications the so - called non - Abelian Stokes theorem (see, for example, \cite{KT}) important for the understanding of confinement, by its Abelian version. This may potentially open new links between the field correlator method \cite{Simonov:2018cbk} and the Abelian projection approach \cite{P}.

The author kindly acknowledges useful discussions with S.Bondarenko, and numerous conversations in the past with M.I.Polikarpov, Yu.A.Simonov and D.I.Diakonov. 

\end{document}